# Ultralow-Temperature Cryogenic Transmission Electron Microscopy Using a New Helium Flow Cryostat Stage

Young-Hoon Kim[1,*], Fehmi Sami Yasin[1,*], Na Yeon Kim[1], Max Birch[2], Xiuzhen Yu[2,3], Akiko Kikkawa[2], Yasujiro Taguchi[2], Jiaqiang Yan[4], Miaofang Chi[1,*]

[1]Center for Nanophase Materials Sciences, Oak Ridge National Laboratory, Oak Ridge, TN 37831, USA

[2]RIKEN Center for Emergent Matter Science (CEMS), Wako, 351-0198, Japan

[3]The Institute of Science Tokyo, Tokyo, 152-8550, Japan

[4]Materials Science and Technology Division, Oak Ridge National Laboratory, Oak Ridge, TN, 37831, USA

[*]Corresponding Authors. *E-mail addresses: kimy6@ornl.gov* (Y.-H. Kim), *yasinfs@ornl.gov* (F. S. Yasin)*, chim@ornl.gov* (M. Chi)

**Abstract**

Advances in cryogenic electron microscopy have opened new avenues for probing quantum phenomena in correlated materials. This study reports the installation and performance of a new side-entry condenZero cryogenic cooling system for JEOL (Scanning) Transmission Electron Microscopes (S/TEM), utilizing compressed liquid helium (LHe) and designed for imaging and spectroscopy at ultra-low temperatures. The system includes an external dewar mounted on a vibration-damping stage and a pressurized, low-noise helium transfer line with a remotely controllable needle valve, ensuring stable and efficient LHe flow with minimal thermal and mechanical noise. Performance evaluation demonstrates a stable base temperature of 6.58 K measured using a Cernox bare chip sensor on the holder with temperature fluctuations within ±0.04 K. Complementary in-situ electron energy-loss spectroscopy (EELS) via aluminum bulk plasmon analysis was used to measure the local specimen temperature and validate cryogenic operation during experiments. The integration of cryogenic cooling with other microscopy techniques, including electron diffraction and Lorentz TEM, was demonstrated by resolving charge density wave (CDW) transitions in $NbSe_2$ using electron diffraction, and imaging nanometric magnetic skyrmions in MnSi via Lorentz TEM. This platform provides reliable cryogenic operation below 7 K, establishing a low-drift route for direct visualization of electronic and magnetic phase transformations in quantum materials.

**1. Introduction**

Recent progress in cryogenic transmission electron microscopy (TEM) has had a major impact on biological and chemical research, culminating in recent Nobel prizes [1–3]. In materials science, cryogenic scanning transmission electron microscopy (S/TEM) is increasingly important for investigating quantum phenomena such as charge density waves, magnetic ordering, superconductivity, ferroelectricity, and topological states etc [4–7]. These collective behaviors often emerge at low temperatures and require advanced imaging techniques for direct observation. The latest generation of microscopy tools offers significantly enhanced spatial, energy, and momentum resolutions, along with magnetic-field-free imaging, transformative capabilities for studying quantum materials [8–12]. However, fully leveraging these advances at atomic resolution demands a cryogenic stage with exceptional mechanical stability, precise temperature control, and flexible tilting capabilities across both ultra-low and intermediate temperature ranges. Such integration has been challenging using previous designs of cryogenic stages. In recent years, new ultra-stable liquid nitrogen ($LN_2$) double-tilt stages have enabled atomic-resolution imaging and spectroscopy at base and intermediate temperatures [13–16]. Nonetheless, liquid helium (LHe) cooling remains essential for exploring quantum phenomena, such as superconductivity, topological states, and correlated electron systems, that often emerge below 100 K.

The development of cooling stages for TEMs dates back to the 1960s, leading to various designs such as superconducting lenses, helium stage modules, top-entry cooling stages, and side-entry cryogenic holders [17–23]. Significant advancements were demonstrated, including Å-level TEM imaging temperatures as low as 1.5 K and energy-loss spectroscopy on solidified gases [22,24,25]. In the 1990s, Gatan Inc. commercialized the side-entry liquid helium cooling holder, based on R. Henderson's design [19], expanding cryo-TEM applications due to its greater accessibility compared to dedicated cryo-microscopes. Atomic-resolution S/TEM imaging and spectroscopy at ~10 K have been demonstrated for studying correlative behavior in quantum materials [26,27]. More recently, the Dresden team replaced the standard objective lens with a large experimental chamber, integrating a continuous-flow liquid helium cryostat into an aberration-corrected microscope. This setup preserves nanometer-scale imaging, enables precise temperature control between 6.5 K and 400 K, and supports long-duration in-situ experiments [28].

Despite these advances, achieving atomic-resolution STEM imaging and energy-loss spectroscopy (EELS) below 100 K remains challenging [26,29]. High-resolution analysis has often relied on luck due to the difficulty in maintaining stage stability in liquid helium experiments. The inherent physical properties of LHe, its tendency to boil and undergo rapid phase transitions, introduce mechanical vibrations and thermal instabilities, both of which compromise the precise conditions required for high-resolution electron microscopy. Additionally, the complexity of cooling and damping systems makes it difficult to incorporate functions such as precise thermal stability at intermediate cryogenic temperatures, beta tilting capabilities, and various in situ capabilities. However, recent innovations from condenZero, H-Bar, NION, and Y. Zhu's team at Brookhaven National Laboratory (BNL) are driving advancements in stable liquid helium-cooled STEM, making atomic-resolution imaging at cryogenic temperatures increasingly viable [29,30].

In this work, we report the first installation and tests of a side-entry LHe-cooling condenZero stage for a JEOL S/TEM, a new flow cryostat featuring an external dewar mounted on a damping stage to minimize and isolate vibrations [31]. We focus on evaluating the cooling performance and quantifying temperature stability and measurement accuracy [32]. The capabilities of this system are further demonstrated by observing charge density wave (CDW) transitions in $NbSe_2$ using electron diffraction and imaging the magnetic skyrmion lattice in MnSi via Lorentz TEM. It is worth noting that a parallel installation of this system for a different type of electron microscope is installed at the ER-C-1 center at Jülich, and its performance may differ from the one reported here. This installation represents the first reported deployment of the condenZero holder system for JEOL microscopes, encompassing utilization of both first- (Gen-1) and second-generation (Gen-2) holder designs, while an improved version is currently under development. Future potential improvements are also discussed in this manuscript.

## 2. Experimental Section/Methods

EELS acquisition of bulk plasmon of Al: To quantify the local specimen temperature at cryogenic conditions, electron energy-loss spectroscopy (EELS) was performed on an electropolished aluminum sample. STEM-EELS measurements were conducted using a Gatan

System (GIF Quantum ER 965, Gatan) in STEM mode at 200 kV accelerating voltage. All the spectrum imaging (SI) datasets were performed at a resolution of 60 × 60 pixels, with a scanning step of 20 nm/pixel, scanning speed of 0.002 s, and energy dispersion of 0.025 eV/channel. The plasmon energy loss spectra were acquired with the zero-loss peak included, and energy alignment was subsequently corrected using the built-in function in the commercial software (Gatan Microscopy Suite, Gatan). To ensure reproducibility, STEM-EELS measurements were performed at least three times under each temperature condition.

Specimen preparation: $NbSe_2$ flakes were prepared by mechanical exfoliation inside a nitrogen-filled glovebox to minimize oxidation. Thin layers were obtained by repeatedly cleaving the bulk crystal with adhesive tape and then transferring the flakes onto TEM support grids. MnSi TEM specimen was prepared from a bulk crystal using a Thermo Fisher Nova 200 Dual Beam FIB, with the ⟨001⟩ axis oriented perpendicular to the thin plate. This orientation served as the TEM optical axis for applying magnetic fields via the objective lens, and the final lamella thickness was measured in the SEM within the FIB system (**Figure S1**).

Lorentz TEM and in-plane magnetic field reconstruction: Lorentz TEM was performed on a JEOL NeoARM S/TEM in low-magnification mode with the objective lens off, producing a nearly field-free sample plane (residual field <9 mT, measured with a Hall sensor). Once stable, we defocused the image and found isolated helical domains hundreds of nanometers in size, as shown in **Figure S2**. These domains appeared and disappeared over time, likely due to beam illumination of adjacent thick regions causing local Joule heating that fluctuated with holder vibrations and electron dose. In thinner regions that are far from the thick edge, no helical domains were observed, consistent with the absence of in-plane ferromagnetic modulations such as a conical state. The micrograph in Figure 6 was acquired as a stack of 100 images (0.081 s exposure each); images with minimal blur were aligned using a custom Python autocorrelation code and integrated to yield an effective exposure of 0.98 s. SITIE reconstruction of this long-exposure image produced the in-plane magnetic induction map shown in Fig. 6a (inset) [33].

## 3. Results and Discussion
### 3.1. Cryogenic electron microscopy

**Figure 1** presents the components of the LHe holder system, developed for ultra-low cryogenic S/TEM. As shown in **Figure 1a**, the system integrates several key components engineered for stable low-temperature operation. To ensure minimal mechanical vibrations, an external LHe dewar is mounted onto a 2.817 m × 0.840 m × 0.840 m permanent fixture damping stage (white box in Figure 1a) which has an active damping system which is calibrated and activated before each experimental session. The helium transfer line (yellow box in Figure 1a) is mounted onto a solid fixture motorized stage capable of lifting/lowering the transfer line independent of the motorized stage used to insert/retract the transfer rod into/from the He dewar. This fixture protrudes an additional 0.559 m × 0.724 m out from the damping stage towards the microscope where it connects directly to the side entry holder, providing a continuous and efficient LHe supply with minimal thermal losses. Its pressurized delivery maintains a steady flow, and insulation reduces heat transfer, minimizing helium boil-off and conserving valuable resources. Precise control of the helium flow rate is achieved through a remotely controllable needle valve, which further enhances vibrational stability. These components are integrated with peripheral equipment, including temperature sensors and control units, allowing for precise thermal regulation and comprehensive operational control. On the back end of the transfer line, the expended He flows through an output line and may be recaptured into He-recovery manifolds. Such a recovery system is currently under construction in our laboratory and will be available in the future.

The sample holder is designed to securely retain the specimen and minimize thermal drift. Detailed views of the holder are presented in **Figures 1b and c**, highlighting its structural design. The system features full JEOL specification compliance with integrated 8-contact biasing functionality. The holder incorporates a high thermal conductivity pathway to efficiently transfer cooling power and minimize temperature fluctuations. **Figure 1c** provides a schematic of the sample loading procedure, which necessitates the use of a specialized chip for the system. The chip accommodates standard TEM grids with a diameter of 3 mm and a thickness of up to 200 μm, which are then secured by an omega-shaped clamp. The cryostat front surface received a gold coating to suppress surface oxidation, and the cross-sectional area connecting the 4 K cold stage to the specimen region was approximately doubled to optimize thermal coupling and cooling capacity. The installation protocol involves (i) securing the specimen chip holding a standard TEM grid with an omega-shaped clamp and insertion into the holder assembly; (ii) establishing electrical connectivity *via* pogo pin contacts by closing the holder tip flap and fastening the small side screw; (iii) mounting the radiation shield around the cryogenic base; (iv) aligning the shield to maintain an unobstructed electron beam path through the aperture and grid center. Insert the sample stage into the operational position after installation and verification. These features collectively minimize charging effects and condensation, ensuring cryogenic conditions for prolonged experimental periods.

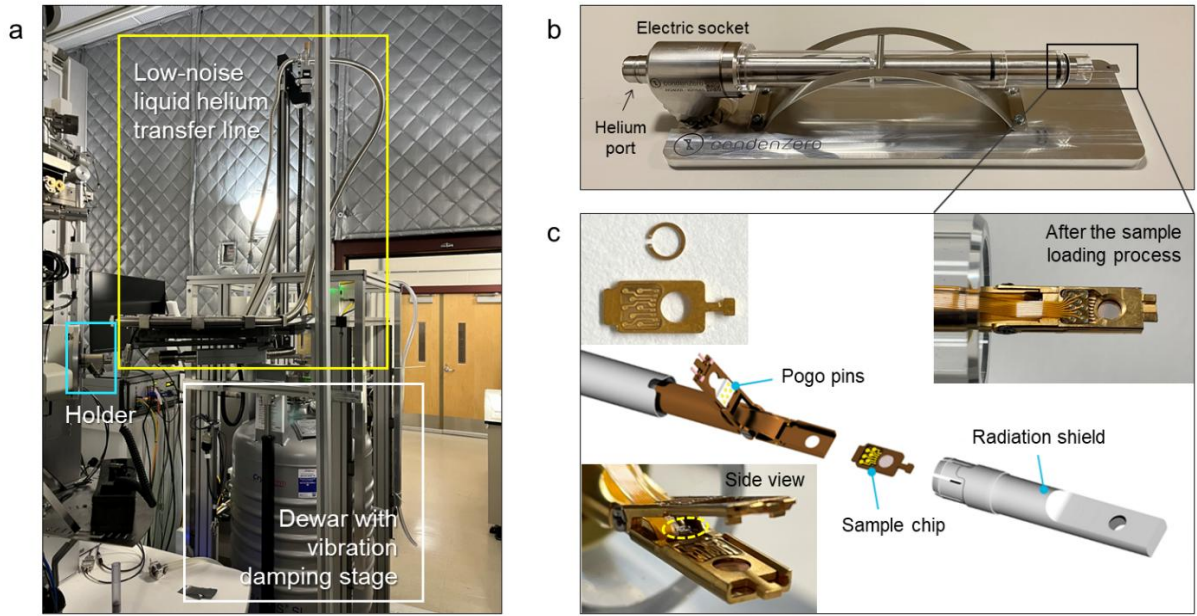

**Figure 1. Components of LHe holder system for cryo-(S)TEM.** (a) Full system configuration, including an external dewar with a damping stage (white box), a helium transfer line (yellow box), a sample holder (cyan box), and peripheral equipment for temperature measurement and operation. (b) Overview of the sample holder. (c) Schematic of the sample loading process, illustrating the use of a specialized chip that accommodates a 3 mm diameter, up to 200 μm thick TEM grid or chip accommodating up to six electrodes. The schematic is reproduced with permission from condenZero [31]. The inset images present the chip and the fully assembled holder following completion of the sample loading process. A radiation shield mounted onto the holder tip after sample loading encloses the assembly for added thermal and mechanical stability. The side view reveals the location of a Cernox bare chip sensor (dashed yellow oval) positioned 5 mm from the specimen at the sample holder base.

Throughout the cryogenic experiment, we continuously monitored the temperature read out from the Cernox bare chip sensor located 5 mm away from the sample at the base of the sample holder flap (at the location labelled indicated by a yellow circle in Figure 1c) to analyze the thermal behavior of the system. **Figure 2a** presents the temperature variations during ultra-low cryo-TEM operation, which are characterized by six distinct stages: (I) room temperature, (II) rapid cooling, (III) slow cooling, (IV) base temperature and (V, VI) warm-up to room temperature. The cooling process (stages I–IV) is primarily governed by the regulated flow of LHe. To further elucidate the temperature dynamics of the cooling process, **Figure 2b** presents a detailed analysis with overlayed plots of the temperature and corresponding dewar pressure. Initially, when the LHe flow begins, a steady pressure rise is observed, followed by a transient surge that briefly exceeds the designated pressure (Stage I). This overshoot, caused by the initial surge in helium flow rate and subsequent system adjustments, stabilizes within approximately 2 min. Before the temperature readout begins decreasing, the LHe cools the sample holder and highly conductive internal components. Once pressure stabilizes and the holder's internal components have sufficiently cooled, the temperature rapidly decreases, reaching approximately 6.1 K (stage II). The process is completed within ~7 min under a stable helium flow at 300 mbar. Subsequently, the temperature continues to decrease gradually until the system reaches its base temperature (stage III). The profile indicates that the base temperature stabilizes at 4.6 K, remaining constant for as long as liquid helium continues to be supplied (stage IV).

To further validate the temperature stability, systematic measurements were performed using both the Gen-1 and Gen-2 holders. As shown in **Figure S3**, magnetic imaging was carried out during 41 min (Gen-1, 100 mbar) and 50 min (Gen-2, 150 mbar) experimental windows marked in the temperature vs. time traces. The corresponding histograms reveal temperature fluctuations of $\sigma = 0.038$ K for the Gen-1 holder and an improved $\sigma = 0.004$ K for the Gen-2 holder (**Figure S3 c, d**). These results confirm that the base temperature of 8.21 K can remain consistently stable over extended periods of time, well suited for long-duration cryogenic TEM experiments. This temperature profile is crucial as thermal perturbations could compromise imaging performance and/or result in fluctuating emergent material properties.

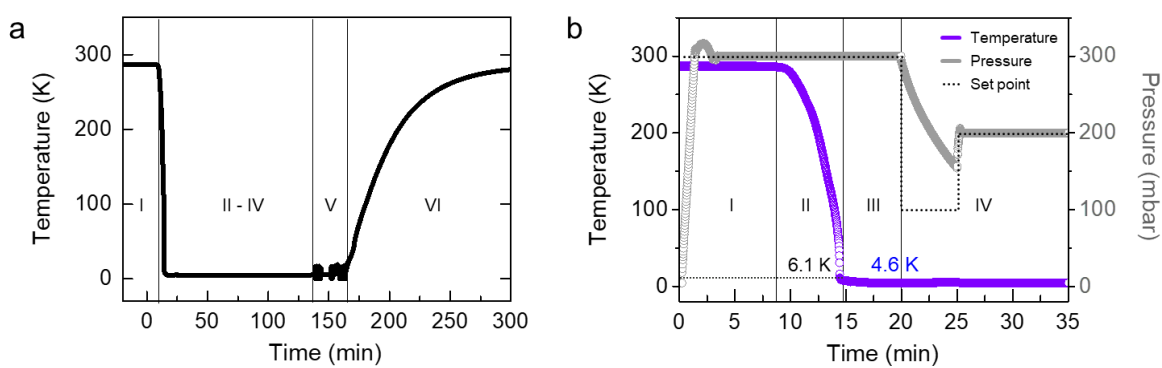

**Figure 2. Temperature variation during cryo-(S)TEM.** (a) Temperature profile consisting of five sequential stages: I – room temperature, II – rapid cooling, III – slow cooling, IV – base temperature, and V, VI – warm-up to room temperature. (b) Magnified temperature profile of stages I–IV with corresponding pressure data. The temperature rapidly drops to 6.1 K within ~7 min, followed by gradual cooling to a base temperature of 4.6 K after another ~6 min. Once the base temperature is reached, the pressure is reduced to the range of 100-200 mbar to enhance imaging stability.

It is important to note that the base temperature can be maintained with minimal temperature fluctuations for many hours, with the only limitation being the dewar size. Each cooldown from room temperature to base temperature was estimated to consume ~10 L of liquid He, and subsequent imaging at base temperature requires approximately 3 L/hr. For a 100 L dewar, this yields a maximum stable imaging time of $\frac{90 \text{ L}}{3 \text{ L/hr}} \approx 30$ hrs, in contrast to the roughly 40-minute window available within side entry He-cooled holders with standalone dewars. This long experimental window is especially important when performing magnetic state mapping experiments or when hunting for exotic spin textures, which usually requires several field sweeps, with alignments at each field value. Following the experiment, the pressure is reduced to zero to eliminate the LHe flow (Stage V), and the system undergoes a passive warm-up phase (Stage VI). Due to the absence of an integrated heating mechanism within the holder, the system naturally returns to room temperature through passive thermal equilibration after helium flow ceases, a process that generally takes slightly over an hour. This gradual thermal recovery ensures safe post-experimental handling of the specimen while preserving system integrity.

To elucidate the interplay between pressure and temperature dynamics, the cooling performance was evaluated as a function of pressure variation. **Figure 3a** presents the temperature profiles at LHe pressures of 700 mbar and 300 mbar. In both cases, an inflection point is observed at 6.1 K, marking the transition between stages II and III. However, at a higher pressure, the cooling process proceeds more rapidly than at lower pressure. At 700 mbar, the temperature dropped to 6.1 K in 185 seconds, followed by an additional 236 seconds to reach the base temperature of 4.4 K. In contrast, at 300 mbar, it took 430 seconds to reach 6.1 K and approximately 400 seconds to reach the base temperature of 4.6 K. These findings confirm that increased LHe pressure not only accelerates the cooling rate but also enables the system to attain lower base temperatures.

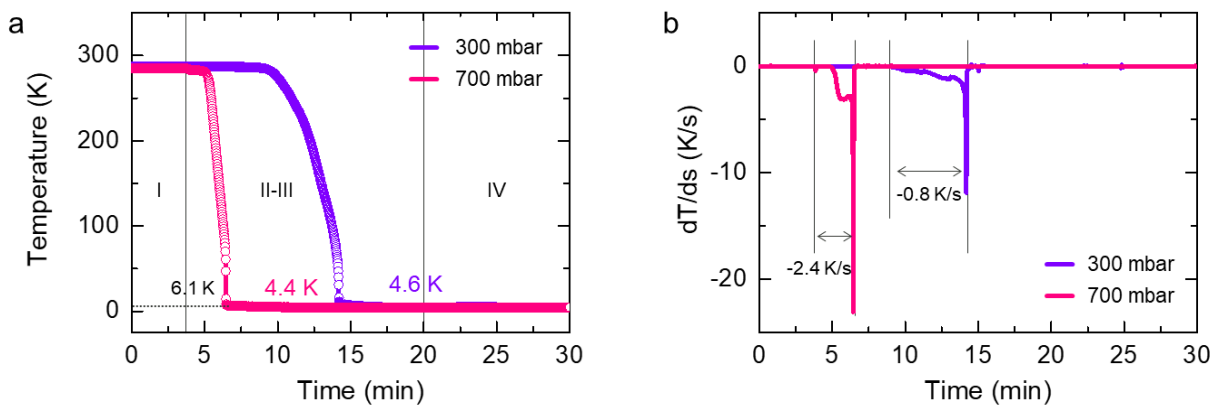

**Figure 3. Evaluation of cooling performance as a function of LHe pressure variation.** (a) Temperature profiles under LHe flow at pressures of $P_{dewar} = 700$ mbar (red) and $P_{dewar} = 300$ mbar (purple). At higher pressure, the system cools more rapidly within ~3 min and reaches a lower base temperature of 4.4 K. The inflection point around 6.1 K marks the transition between rapid and slow cooling phases (Stages II and III). (b) First derivative of the temperature as a function of time $\frac{dT}{dP_{dewar}}$ for $P_{dewar} = 700$ mbar plotted in red and $P_{dewar} = 300$ mbar plotted in purple.

**Figure 3b** plots the first derivative of the temperature as a function of time, which provides quantitative insight into the cooling dynamics under different pressure conditions. The most pronounced rate of temperature change is observed during the rapid cooling stage (Stage II), with the maximum derivative reaching –23.1 K/s at 700 mbar and –11.9 K/s at 300 mbar. By averaging the rate of temperature change over stage II, we determined the mean cooling rates to be approximately 2.4 K/s at 700 mbar and 0.8 K/s at 300 mbar. Upon transitioning to Stage III, the temperature change converges to approximately 6 mK/s and 4 mK/s for each condition, indicating a gradual and uniform approach toward thermal equilibrium. These results underscore the critical role of LHe flow control in optimizing cryogenic performance. While elevated flow enables faster and deeper cooling, the convergence of cooling rates in Stage III suggests that a finite amount of time is still required for effective thermal transfer to the holder tip. Therefore, developing optimized cooling protocols is essential—not only to achieve low temperatures rapidly but also to ensure sufficient stabilization time for uniform thermal distribution across the holder tip. The results shown here correspond to the second generation of holder design, whereas the results of the previous holder are included in the supporting information (**Figures S4 and S5**).

*3.2. Evaluation of local specimen temperatures*

The sample holder is equipped with an integrated thermometer located on the rear side of the specimen mounting region, enabling real-time monitoring of temperature fluctuations. This built-in thermometer uses a calibrated Cernox sensor and is read out via a four-point resistivity measurement. To assess the actual local temperature of the specimen, electron energy loss spectroscopy (EELS) was employed, and the results were compared with those obtained from the built-in thermometer. The thermal response of aluminum—thermal expansion at elevated temperatures and contraction upon cooling—directly influences its internal electron density. As the temperature decreases, lattice contraction leads to an increase in valence electron density, which in turn induces an energy shift (ΔE) in the bulk plasmon [34]. To quantify this behavior, the temperature-dependent thermal expansion coefficient of aluminum, denoted as $\alpha_l(T)$, was extracted from previously reported data [35]. The coefficient ($\alpha_l$) was fitted using a fourth-order polynomial function, and the resulting fitting constants, along with the fitted curve, are presented in **Figure S6**. Subsequently, the coefficient ($\alpha_l$) was integrated to calculate $f(T)$. This function was then applied to the free-electron model to determine the temperature-dependent bulk plasmon energy ($E_p(T)$) using the relation $E_p(T) \approx E_p(T_0)[1 - \frac{3}{2}f(T)]$. The reference value for the bulk plasmon energy was ~15.23 eV at the reference temperature ($T_0$ = 300 K), consistent with previous studies [32,36].

To investigate the temperature-dependent behavior of the bulk plasmon, a series of spectra were acquired from electropolished aluminum at various temperatures ranging from room temperature to the base temperature, as indicated by the thermometer (**Figure 4a**). During the experiment, LHe was continuously supplied to cool the sample to base temperature (10 K in this case), after which the dewar pressure and therefore LHe flow was reduced to allow the temperature to gradually increase. A systematic blueshift of the plasmon peak was observed with decreasing temperature, reflecting the temperature-dependent contraction of the aluminum lattice. To ensure reproducibility, STEM-EELS measurements were performed at least three times under each temperature condition, and the bulk plasmon peak positions were statistically quantified. **Figure 4b** presents the peak positions extracted from the EELS spectra as a function of temperature, where the x-axis corresponds to the temperatures recorded by the built-in thermometer. The red line represents the theoretically derived bulk plasmon energies based on the linear thermal expansion coefficient of aluminum. The experimentally quantified peak positions include 15.253 eV at 260 K, 15.276 eV at 205 K, 15.319 eV at 155 K, 15.309 eV at 90 K, 15.293 eV at 65 K and 15.324 eV at 43 K.

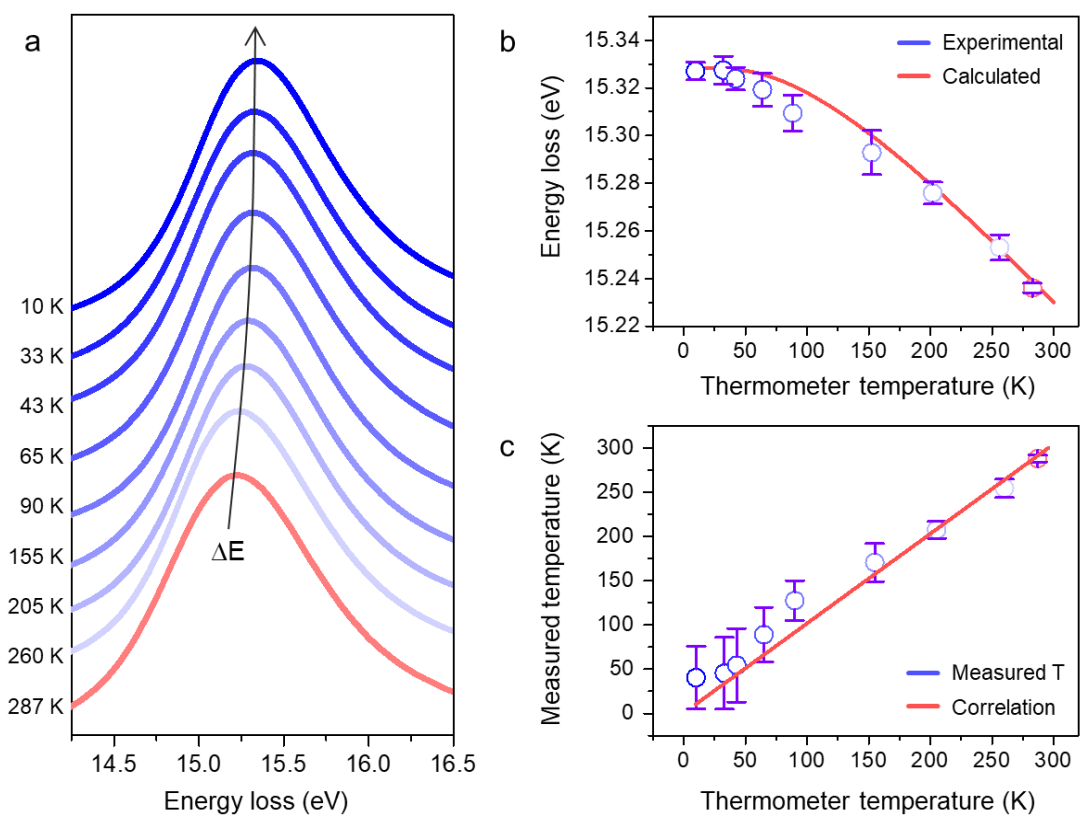

**Figure 4. Evaluation of local specimen temperature via EELS-based plasmon energy shift in aluminum.** (a) Representative bulk plasmon spectra obtained from electropolished aluminum at various temperatures ranging from 287 K to 10 K, as indicated by the thermometer. A systematic blueshift of the plasmon peak is observed as temperature decreases, reflecting lattice contraction. (b) Extracted bulk plasmon peak energies plotted as a function of temperature. The red line corresponds to theoretically calculated values based on the thermal expansion coefficient of aluminum, while the blue dots indicate experimentally measured peak positions. (c) Comparison between local specimen temperatures estimated from EELS and those recorded by the built-in thermometer.

As a result, pointwise deviations between the EELS-derived temperatures and the thermometer readings (**Figure 4c**) were ~12 K at 33–43 K, ~24 K at 65 K, and ~37.5 K at 90 K. Above 35 K, this method shows good agreement with the readings from a calibrated Cernox sensor, with only minor discrepancies that can be attributed to differences in measurement location: the plasmon shift reflects the temperature at the electron probe position, whereas the Cernox sensor is mounted ~5 mm away on the base of the cryostat. Despite these local discrepancies, the average deviation across the entire temperature range remained within 15.6 K. These differences are likely due to the measurement procedure, where EELS was collected during a passive warm-up phase with minimal LHe flow. In other words, in this regime, where no active heating is applied and the temperature is governed by local thermal equilibrium between residual LHe flow and ambient conditions, relatively larger discrepancies may arise in temperature intervals exhibiting more dynamic thermal transitions (e.g., around 65 K and 90 K) compared to other temperature intervals. Note that the specimen temperature may also increase at the specimen due to the electron flux during imaging. Beam-induced heating should be quite small for the metallic samples studied here. Previous studies estimate that the beam-induced heating in polymers is likely limited to approximately 1 K for beam currents of 100 pA [37].

Below 35 K, the thermal expansion of Al changes very little with temperature, meaning that even small shifts in plasmon energy can correspond to relatively large variations in actual temperature, leading to the apparent mismatch between EELS and thermometer readings (**Figure 4c**). More accurate methods for local temperature calibration at the specimen are therefore needed. One approach is to use phase-transition materials, which can provide precise calibration in this regime. However, thin TEM specimens, particularly two-dimensional materials, may not exhibit the same transition temperature (Tc or $T_N$) as their bulk counterparts. Therefore, special caution is required when using phase transitions for calibration in cryogenic TEM. Another approach is to integrate resistance sensors, for example, utilizing Micro-Electro-Mechanical Systems (MEMS)-based chips, which in principle, enable reliable local temperature measurements at cryogenic conditions. Achieving this will require future designs with robust MEMS chips and low-resistance contacts to ensure mechanical stability while minimizing strain on the specimen during rapid cooling to ultralow temperatures. Regardless, our findings confirm the thermometer readings reflect the actual specimen temperature and validate the reliability of the temperature monitoring system integrated into the holder. The ability to experimentally determine the local specimen temperature via EELS also provides a valuable cross-validation tool, and the demonstrated thermal stability of this cryogenic holder facilitates the implementation of a robust and efficient cryogenic environment for future experiments.

### 3.3. NbSe$_2$ with charge density wave ordering

To assess the diffractive imaging performance of the LHe-based cryogenic TEM system, NbSe$_2$ was selected as a model layered quantum material. The thermodynamically stable form under ambient conditions, 2H-NbSe$_2$, adopts a hexagonal structure (space group P6$_3$/mmc) consisting of layered niobium atoms coordinated trigonal prismatically by six selenium atoms (**Figure 5a**). This material is well known for undergoing a charge density wave (CDW) phase transition at approximately 33 K, followed by superconductivity below 7 K [38]. Notably, the CDW phase forms a structural modulation (**Figure 5b**), inducing a periodic lattice distortion that breaks the ideal hexagonal symmetry by clustering Nb atoms into 3×3 supercells [38,39].

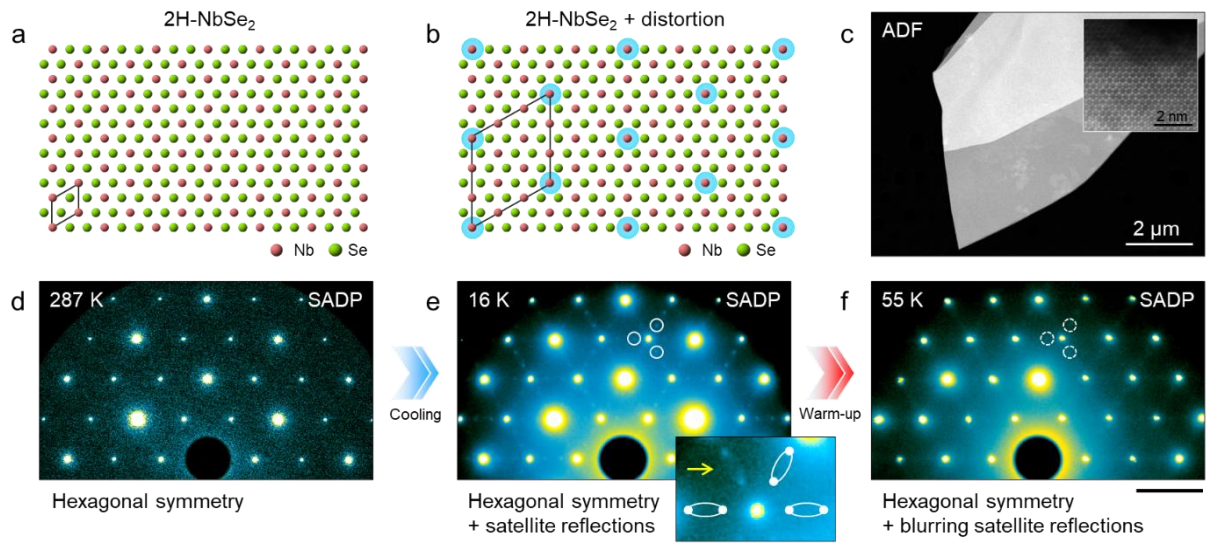

**Figure 5. Cryogenic electron diffraction analysis of CDW modulation in NbSe$_2$.** (a) Crystal structure of 2H-NbSe$_2$, showing the hexagonal layered configuration with niobium atoms coordinated trigonal prismatically by selenium atoms. (b) Schematic illustration of the superlattice of CDW ordering. (c) ADF image of a mechanically exfoliated NbSe$_2$ flake. (d–f) selected area electron diffraction patterns (SADP) acquired at room temperature, 16 K, and 55 K, respectively. At 16 K, additional satellite reflections appear at 1/3 positions in reciprocal space, providing evidence of CDW ordering. At 55 K, a clear weakening of these satellite peaks is observed, indicating the thermal suppression of the modulation for CDW. Scale bar is 5 1/nm.

To probe this structural modulation, a mechanically exfoliated NbSe$_2$ flake (**Figure 5c**) was transferred onto a TEM grid for cryogenic electron diffraction analysis. The NbSe$_2$ single crystals used in this study were grown via the chemical vapor transport method using I$_2$ as the transport agent. NbSe$_2$ powder was placed at the hot end of a sealed ampoule at 800 °C, while plate-like single crystals formed at the cold end maintained at 750 °C. At room temperature, the selected area electron diffraction pattern (SADP) exhibited only the Bragg reflections (**Figure 5d**), consistent with the pristine hexagonal lattice symmetry of 2H-NbSe$_2$. However, upon cooling below 33 K, additional satellite reflections emerged at positions corresponding to 1/3 of the reciprocal lattice vectors (**Figure 5e**). The appearance of these reflections at 16 K reflects a direct evidence for periodic lattice modulation associated with superlattice formation for the charge density wave (CDW) ordering, likely attributed to periodic displacements that break the translational symmetry of the original lattice [38,39]. Interestingly, the satellite peaks follow a honeycomb-like arrangement but appear to exhibit directional dependence. In addition, elliptical contours are observed surrounding these satellite reflections in reciprocal space, which may be indicative of anisotropic domain structures, local strain fields, or incomplete coherence in the CDW modulation. Further exploration of the satellite peak morphology, including its potential dependence on the cooling rate, are planned to provide valuable insights into the dynamics of domain formation and the coherence of CDW ordering. When the flow of LHe was halted, and the specimen temperature was allowed to rise, the SADP obtained at 55 K revealed a noticeable weakening of the superlattice peaks (**Figure 5f**). This phenomenon indicates the attenuation of CDW modulation as a function of increasing temperature. These findings demonstrate not only the evolution of the CDW phenomenon in NbSe$_2$ but also the capability and effectiveness of the cryogenic S/TEM system to capture subtle structural transitions with fidelity to ultra-low cryogenic temperature for quantum materials.

*3.4. Magnetic imaging of nanomagnetic textures in MnSi*

To benchmark the imaging stability of the LHe system in real space, we utilized MnSi, a prototypical chiral cubic helimagnet with a Curie temperature of $T_C \approx 22.5$ K in nanofabricated thin plates (compared to $T_C = 29.5$ K in bulk crystals) and known for its magnetic skyrmion lattice emergent at elevated applied magnetic fields [40–42]. Spin textures in this system exhibit a helical wavelength of $\lambda_{hel} = 18$ nm in nanowires and nanofabricated thin plates, which matches the skyrmion lattice constant [41,42] and is several times larger than the resolution achievable with Lorentz S/TEM, in which the objective lens is adjusted to create a field-free sample environment.

Here, we employed a Ga-ion focused ion beam (FIB) milling technique (see the Methods section for details) to fabricate a MnSi thin plate with the ⟨001⟩ crystal axis oriented normal to the plate. After thinning the plate to 144 nm, we inserted the sample into the LHe holder and using a dewar pressure $P_{dewar} > 700$ mbar, cooled it to 6.58 K, as indicated by the temperature controller and depicted in the temperature profile shown in **Figure S3**. Using Lorentz TEM mode, we initially observed significant lateral sample stage vibrations at this dewar pressure and proceeded to decrease the pressure to $P_{dewar} = 100$ mbar, after which the lateral sample translations decreased enough to perform stable imaging. At this decreased dewar pressure, the measured temperature increased to $T = 8.21 \pm 0.038$ K and remained stable at this temperature for the duration of the experiment, approximately 41 minutes.

3.5. *Evaluating* stage stability *via sequential MnSi images*

To evaluate the mechanical stability, we acquired a temporal imaging sequence at the base temperature of 5.6 K under a dewar pressure of 110 mbar (**Figure S7**). The analysis indicates that the lateral perturbations exhibited amplitudes of $\Delta_X = 0.80 \pm 0.53$ nm/s and $\Delta_Y = 0.77 \pm 0.42$ nm/s. By decomposing the signal into high-frequency components and low-frequency drift constituents, the vibrational amplitudes were determined using root mean square RMS as X = 1.0 nm and Y = 1.2 nm, and the drift was quantified as $\Delta_X = 0.74 \pm 0.53$ nm/s and $\Delta_Y = 0.77 \pm 0.41$ nm/s along each direction (**Figure 6a, b**). In contrast, measurements conducted under elevated pressure conditions at 4.4 K under pressure of 800 mbar revealed relatively large mechanical instabilities (**Figure 6c, d**). The vibrational amplitudes increased to X = 4.3 nm and Y = 2.8 nm, with the drift components of $\Delta_X = 4.01 \pm 2.38$ nm/s and $\Delta_Y = 0.02 \pm 0.09$ nm/s. It should be noted that this stability is not sufficient for atomic-scale imaging, but it is adequate to enable nanoscale cryogenic Lorentz magnetic imaging through the implementation of fast-rate, multi-frame imaging using high-speed direct electron cameras. Furthermore, the vendor is continuing to improve the holders to enhance their mechanical stability, and future generations are expected to further expand the range of high-resolution cryogenic imaging applications.

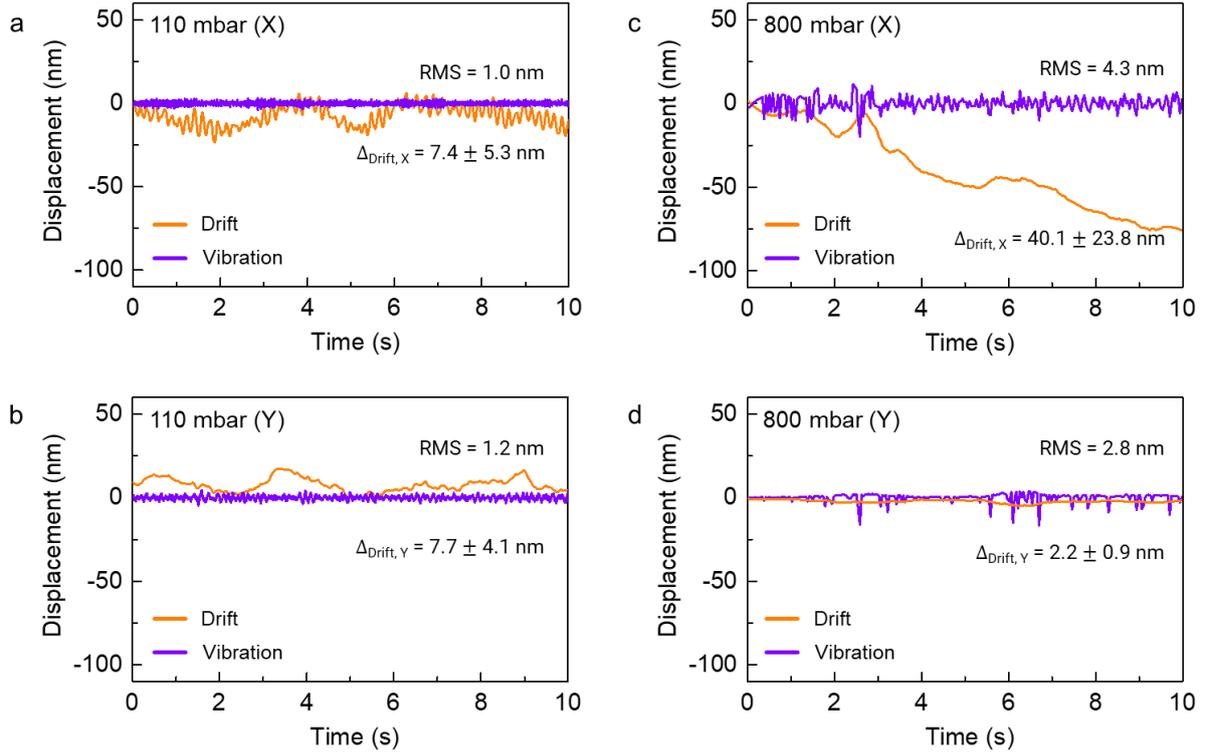

**Figure 6. Comparative analysis of specimen motion decomposed into high-frequency vibrations and low-frequency drift components under two distinct pressure conditions at base temperatures.** (a, b) At reduced pressure condition (110 mbar, T = 5.6 K), vibration amplitudes expressed as root mean square (RMS) values were determined to be 1.0 nm (X) and 1.2 nm (Y), with drift measurements of $\Delta_X = 7.4 \pm 5.3$ nm and $\Delta_Y = 7.7 \pm 4.1$ nm over 10 s temporal interval. (c, d) At elevated pressure conditions (800 mbar, T = 4.4 K), large vibrational perturbations were observed with RMS amplitudes of 4.3 nm (X) and 2.8 nm (Y), accompanied by significantly increased drift values of $\Delta_X = 40.1 \pm 23.8$ nm and $\Delta_Y = 2.2 \pm 0.9$ nm over the equivalent acquisition period. Operation under reduced pressure suppresses both high-frequency mechanical oscillations and low-frequency positional instabilities.

To create a magnetic skyrmion lattice state, we excited the objective lens current to induce a magnetic field $\mu_0 H_{ext}$ along the ⟨001⟩ crystal axis. As shown in **Figure 7**, a skyrmion lattice emerged at $\mu_0 H_{ext} = 324$ mT with lattice spacing $a_{sky} = 18.3 \pm 0.2$ nm, in good agreement with previous reports [40–42]. The six-fold lattice symmetry is displayed in the LTEM micrograph in **Figure 7a** and is shown clearly in the discrete Fourier transform displayed in supplementary **Figure S8**. The vortex-like real-space in-plane structure is shown in the in-plane magnetic induction map in **Figure 7b** calculated using the single image transport of intensity equation (SITIE) [33]. Furthermore, the mechanical stability of the holder coupled with a high-speed detector allowing for long 0.625 exposures after drift correction results in a sufficient signal-to-noise ratio to resolve the magnetic domain grain boundaries separating skyrmion lattice domains with relative twist angles. Such a grain boundary is shown in **Figure 7c-d**, separating two domains with a twist angle of 6°.

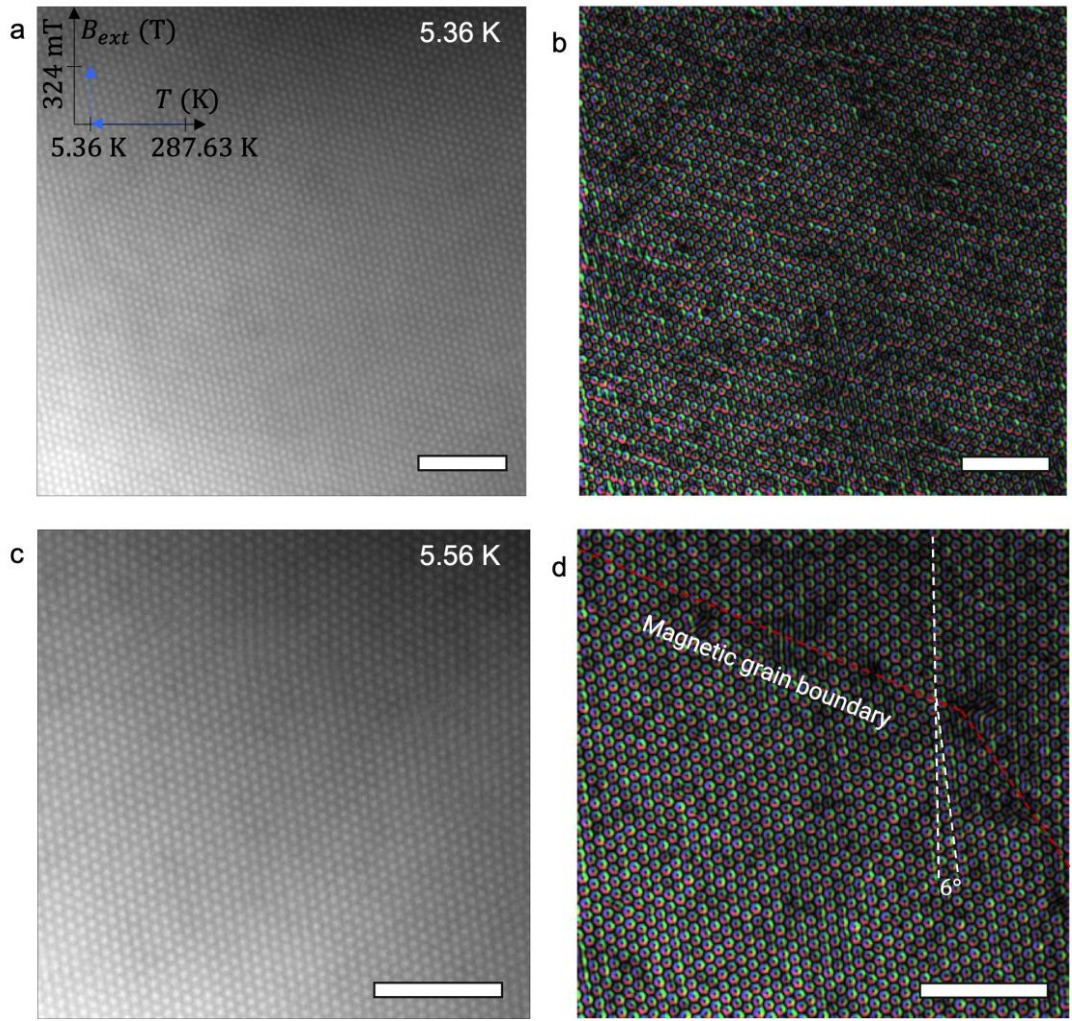

**Figure 7. Magnetic imaging of nanomagnetic textures in MnSi.** (a,c) Lorentz transmission electron micrograph of the hexagonal skyrmion lattice emergent in a MnSi lamellae with defocus $\Delta f = -66$ μm, 0.625 s exposure time, and applied external magnetic field $\mu_0 H_{ext} = $ 324 mT at 5.36 K and 5.56 K, respectively. Scale bars 200 nm. (b, d) Corresponding in-plane magnetic induction maps reconstructed using the single image transport-of-intensity equation (SITIE). The magnetic domain grain boundary marked by the dash red line separates two skyrmion lattice domains rotated with respect to one another by 6°. Scale bars 200 nm.

The magnetic skyrmion lattice state permeated the entire field of view in a thin region of the sample where the zero field spin textures were invisible, allowing us to focus the direct beam so that no electrons illuminated the thick sample region near the TEM mesh grid. While the micrographs in **Figure 7** were acquired via drift correction of ≈ 353 frames over a 0.625 s exposure, we also obtained the LTEM micrographs shown in **Figure S8** by acquiring a stack of 100 images with a 12.35 frame-per-s frame rate. We registered, aligned and integrated the frames unaffected by lateral translations to obtain a final image with an accumulated exposure time of 0.98 s, demonstrating the potential of this holder for use with both conventional and fast electron detectors. Furthermore, we observed that the skyrmion lattice began polarizing into a ferromagnetic state above $\mu_0 H_{ext}$ = 350 mT, which is on the same order of magnitude as previous real-space studies [41,42]. Our results confirm that this holder provides stable, long-time imaging experiments of magnetic materials with spin ordering temperatures well below liquid nitrogen temperatures.

## 4. Conclusions

This study presents a characterization of the performance of a liquid helium (LHe) flow-based cryogenic sample holder system integrated into a JEOL scanning/transmission electron microscope (S/TEM). The system, designed with an externally mounted dewar and a pressurized helium transfer line, achieves a base temperature of 6.58 K with exceptional thermal stability, maintaining temperature deviations within a narrow range of ±0.04 K. The temperature measurement was validated through local EELS-based calibration using plasmon peak shifts in aluminum. The capabilities of this system for studying quantum phase transitions were further demonstrated through two examples: electron diffraction revealed the 3×3 charge density wave (CDW) ordering in $NbSe_2$ at 16 K, and Lorentz TEM imaging of MnSi at 8.2 K confirmed the formation of a nanometric magnetic skyrmion lattice under an applied magnetic field. These results highlight the power of integrating advanced microscopy techniques at LHe temperatures to directly visualize quantum phenomena in materials. Looking ahead, we anticipate that the next generation of condenZero stages and other emerging LHe cooling systems, both side-entry and fully integrated, will offer improved stage stability at sub-10 K and intermediate temperatures. Furthermore, this holder boasts six electrodes, which will be used for a range of *in situ* measurements, including uniform heating and electric and thermal current-driven dynamic phenomena. These improvements will open the door to promising real-space investigations of quantum phase transitions, nonequilibrium dynamics, and emergent excitations in correlated quantum materials.

**Acknowledgements**

This work was supported by the U.S. Department of Energy, Office of Basic Energy Sciences, Division of Materials Sciences and Engineering under contract FWP-ERKCS89. Microscopy technique development (Y.-H.K.) was partially supported by the U.S. BES, Early Career Research Program (KC040304-ERKCZ55). Experiments were performed at the Center for Nanophase Materials Sciences, a U.S. DOE Office of Science User Facility at Oak Ridge National Laboratory (ORNL). Research sponsored by the Laboratory Directed Research and Development Program of ORNL, managed by UT-Battelle, LLC, for the US DOE. X. Z. Y and Y. T acknowledge the support of the Japan Science and Technology Agency (JST) CREST program (# JPMJCR20T1) and the RIKEN TRIP initiative. The authors thank CondenZero GmbH for their support and valuable discussions.

**Declaration of Competing Interest**

The authors declare no competing interests.

**Author Contributions**

Y.-H.K., F.S.Y. and M.C. conceived this work, wrote, and edited the manuscript. Y.-H.K and F.S.Y. evaluated the cooling performance and mechanical stability. J.Y. grew the bulk crystal of $NbSe_2$. Y-H.K. and N.K. performed mechanical exfoliation and transferred it onto a TEM grid. Y.-H.K. performed cryogenic EELS of bulk plasmon and SADP measurements for $NbSe_2$ with charge density wave ordering. A.K. and Y.T. grew the MnSi crystal, and M.B., X.Z.Y., and F.S.Y. prepared the MnSi specimen. F.S.Y. performed the cryogenic real-space imaging. All authors participated in the manuscript review.

**Data Availability Statement**

Data will be made available on request.

Supporting Information

# Ultralow-Temperature Cryogenic Transmission Electron Microscopy Using a New Helium Flow Cryostat Stage

Young-Hoon Kim[1,*], Fehmi Sami Yasin[1,*], Na Yeon Kim[1], Max Birch[2], Xiuzhen Yu[2,3], Akiko Kikkawa[2], Yasujiro Taguchi[2], Jiaqiang Yan[4], Miaofang Chi[1,*]

[1]Center for Nanophase Materials Sciences, Oak Ridge National Laboratory, Oak Ridge, TN 37831, USA
[2]RIKEN Center for Emergent Matter Science (CEMS), Wako, 351-0198, Japan
[3]The Institute of Science Tokyo, Tokyo, 152-8550, Japan
[4]Materials Science and Technology Division, Oak Ridge National Laboratory, Oak Ridge, TN, 37831, USA

*Corresponding Authors. *E-mail addresses:* kimy6@ornl.gov (Y.-H. Kim), yasinfs@ornl.gov (F. S. Yasin)*, chim@ornl.gov* (M. Chi)

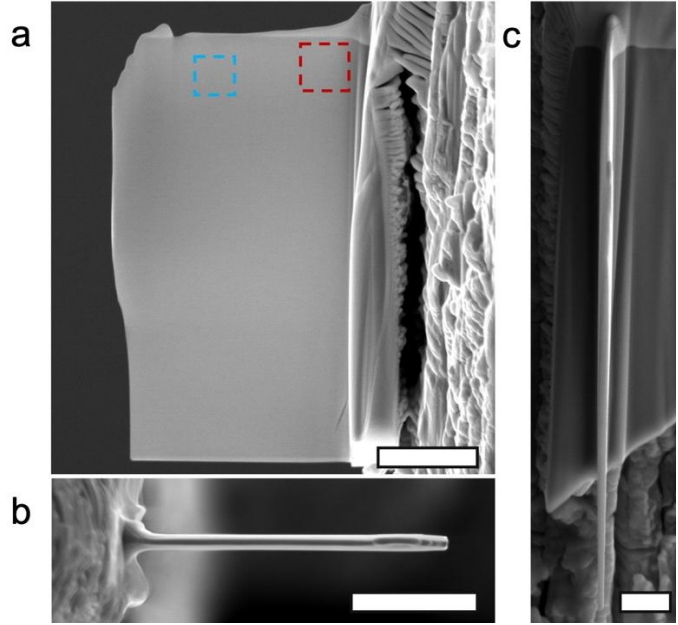

**Figure S1.** Scanning electron micrographs of the MnSi lamella fabricated using Ga-ion FIB. (a) Front profile of the thin plate taken with a stage tilt of 52°. The approximate fields of view where we performed real space imaging in Figure 6 and Figure S2 are indicated by the dashed blue and red squares, respectively. Scale bar 2 µm. (b) Top-down profile of the thin plate, indicating a thickness of $d \approx 144$ nm near the top edge. Scale bar 2 µm. (c) Side profile of the thin plate, indicating a relatively uniform thickness near the top edge, with a linearly decreasing thickness near the bottom edge. Scale bar 1 µm.

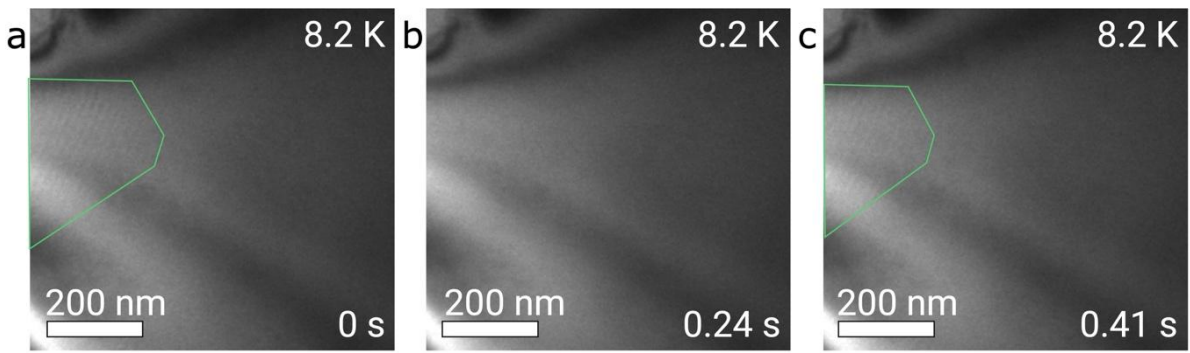

**Figure S2.** Magnetic imaging of nanomagnetic textures in MnSi. (a) Lorentz transmission electron micrograph of the initial magnetic state in a MnSi lamellae with defocus $\Delta f = 60$ µm and applied external magnetic field $\mu_0 H_{ext} = 0$ mT. Helical stripe domains hundreds of nanometers in size are present but periodically disappear and reappear over time.

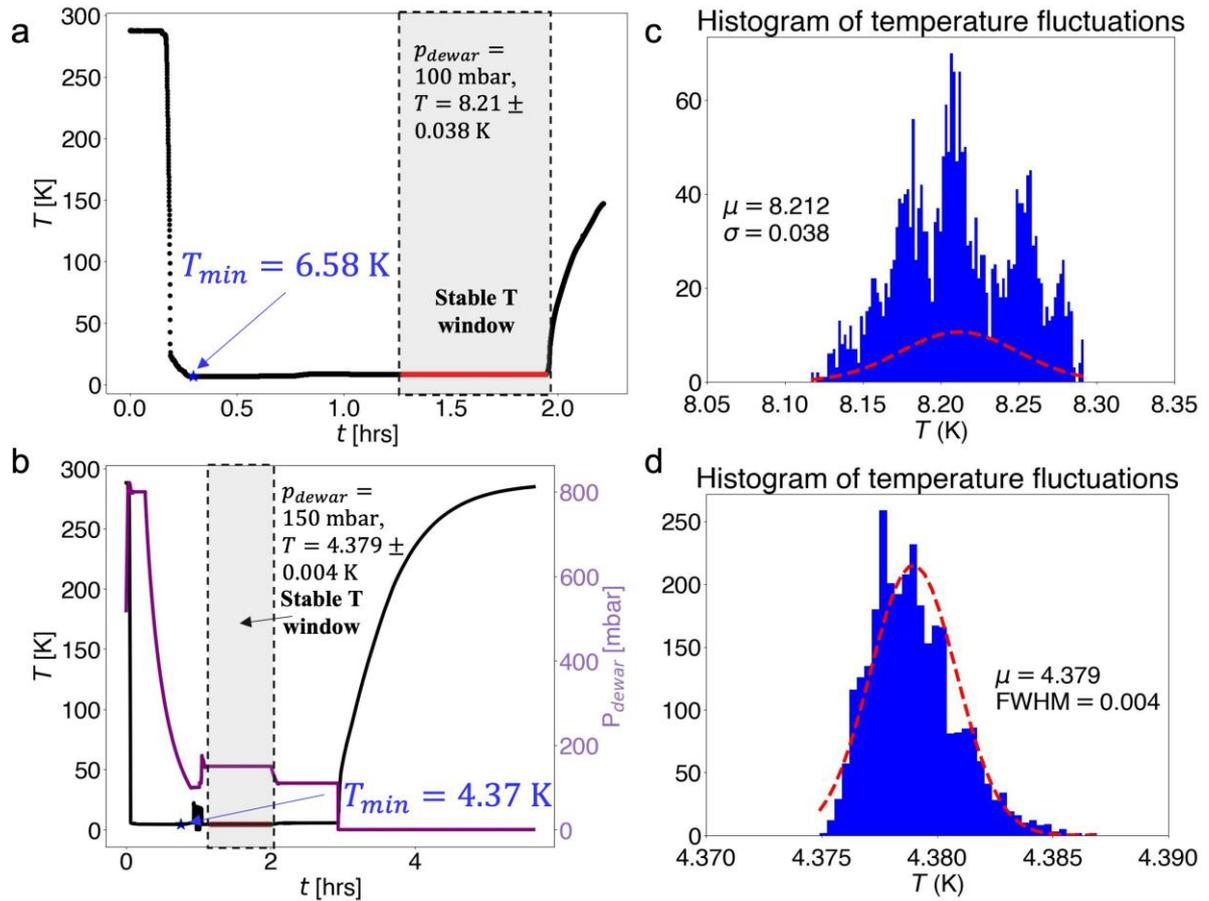

**Figure S3**. (a, b) Temperature profiles during LHe flow for (a) the initially installed holder (Gen-1) and (b) the upgraded holder (Gen-2). We performed magnetic imaging during the 41 min and 46.5 min time regions indicated by the red markers at dewar pressures of 100 mbar and 150 mbar, respectively. (c, d) Histogram of temperature fluctuations during the magnetic imaging experiments demonstrating temperature fluctuations of σ = 0.038 K and 0.004 K, respectively.

**Note S1. Temperature stability during real-space imaging of the magnetic skyrmion lattice state in a nanofabricated MnSi thin plate.**

**Figure S1** shows selected scanning electron microscopy (SEM) images of the nanofabricated MnSi thin plate, with front surface, top-down, and side profiles displayed in Figure S1a, b and c, respectively. We used the top-down profile in Figure S1b to estimate the thickness near the top edge of the thin plate to be $d \approx 144$ nm, which is the edge near the field of view we used to perform the real-space imaging shown in Figure 6, which is indicated by the dashed blue square overlayed on Figure S1a.

When we initially began imaging the MnSi thin plate at base temperature, we observed helical domains hundreds of nanometers in size that came in and out of contrast as a function of time. This behavior is displayed in the Lorentz TEM micrographs shown in **Figure S2a-c** taken at time 0 s, 0.24 s, and 0.41 s, respectively. This may have been due to the edge of the electron beam illuminating the thicker region ($> 2$ μm) of the sample adjacent to the field of view in Figure S2, with these electrons being absorbed by the sample and flowing to the mesh grid, causing local Joule heating that fluctuates as the sample holder lateral vibrations results in fluctuations in the electron dose at the thick region and therefore a fluctuating temperature. When we surveyed the thin sample region sufficiently far away from the thick region so that the direct beam did not illuminate it, we did not observe helical domains, suggesting the absence of a ferromagnetic state with modulations oriented in the plane of the thin plate. Such a lack of contrast could be due to a helical state or states or even a conical state with a modulation vector oriented out-of-plane. Alternatively, the helices may be depinned and dynamically translating or rotating across the field of view too rapidly to be captured by our detector's frame rate, which was nominally 0.08 s.

**Figure S3a** shows the temperature as a function of time from cool down of the sample, to stable real-space imaging performed at $T = 8.21 \pm 0.038$ K indicated by red markers, to the final warm-up two hours later. As described in the main text, we decreased the dewar pressure to 100 mbar to achieve a holder stability good enough to perform the nanometric resolution real-space magnetic imaging seen in Figure 6. At this decreased dewar pressure, the temperature fluctuations were within a standard deviation of 38 mK from the mean temperature of 8.21 K, as shown in **Figure S3b**. At this pressure, mechanical vibrations were minimal, making it the most stable operating point. At lower pressures (<80 mbar), the LHe flow becomes too weak, likely to lead to intermittent bubble formation and collapse in the transfer line, which introduces pressure pulses that couple into the holder as vibrations. At higher pressures, helium flows too rapidly through the transfer line, possibly generating turbulence and continuous bubbling, which likewise induces mechanical vibrations and acoustic noise.

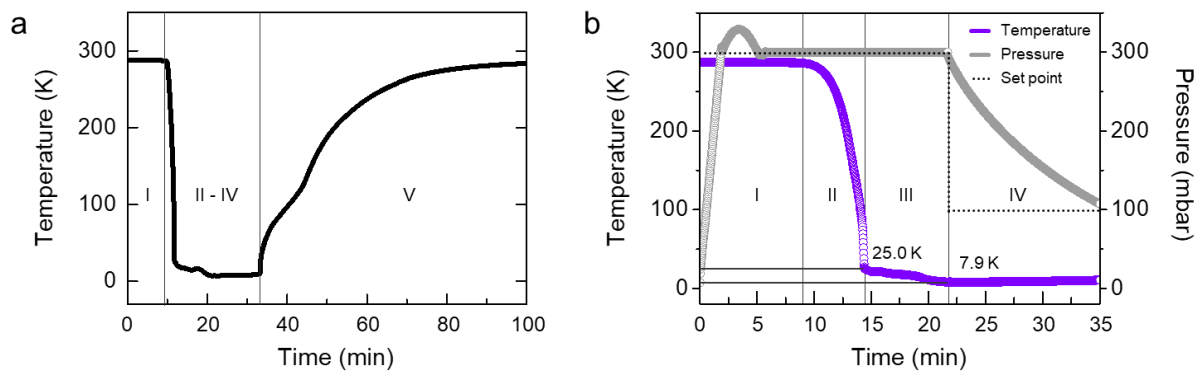

**Figure S4. Temperature variation during cryo-(S)TEM.** (a) Temperature profile consisting of five sequential stages: I – room temperature, II – rapid cooling, III – slow cooling, IV – base temperature, and V – warm-up to room temperature. (b) Magnified temperature profile of stages I–IV with corresponding pressure data. The temperature rapidly drops to 25 K within ~7 min, followed by gradual cooling to a base temperature of 7.9 K after another ~8 min. Once the base temperature is reached, the pressure is reduced to 100 mbar to enhance imaging stability.

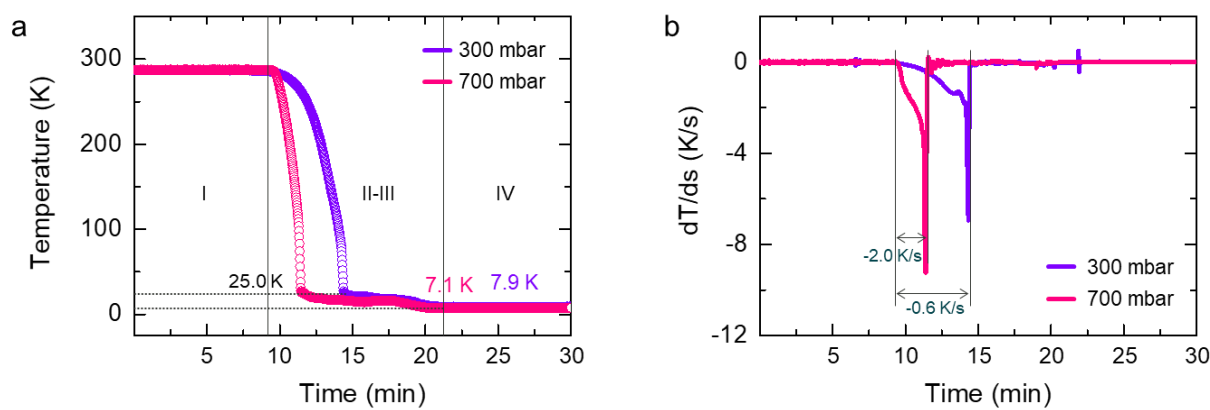

**Figure S5. Evaluation of cooling performance as a function of LHe pressure variation.** (a) Temperature profiles under LHe flow at pressures of $P_{dewar} = 700$ mbar (red) and $P_{dewar} = 300$ mbar (purple). At higher pressure, the system cools more rapidly within ~2 min and reaches a lower base temperature of 7.1 K. The inflection point around 25 K marks the transition between rapid and slow cooling phases (Stages II and III). (b) First derivative of the temperature as a function of time $\frac{dT}{dP_{dewar}}$ for $P_{dewar} = 700$ mbar plotted in red and $P_{dewar} = 300$ mbar plotted in purple.

**Note S2. Temperature variation during cryo-(S)TEM using Gen-1 holder**

**Figure S4a** presents the temperature variations during ultra-low cryo-TEM operation, which are characterized by five distinct stages: (I) room temperature, (II) rapid cooling, (III) slow cooling, (IV) base temperature and (V) warm-up to room temperature. The cooling process (stages I–IV) is primarily governed by the regulated flow of LHe. **Figure S4b** presents a detailed analysis with overlayed plots of the temperature and corresponding dewar pressure. Initially, when the LHe flow begins, a steady pressure rise is observed, followed by a transient surge that briefly exceeds the designated pressure (Stage I). This overshoot, caused by the initial surge in helium flow rate and subsequent system adjustments, stabilizes within approximately 3 min. Before the temperature readout begins decreasing, the LHe cools the sample holder and highly conductive internal components. Once pressure stabilizes and the holder's internal components have sufficiently cooled, the temperature rapidly decreases, reaching approximately 25 K (stage II). The process is completed within ~7 min under a stable helium flow at 300 mbar. Subsequently, the temperature continues to decrease gradually until the system reaches its base temperature (stage III). The profile indicates that the base temperature stabilizes at 7.9 K, remaining constant for as long as liquid helium continues to be supplied (stage IV). To ensure optimal real-space imaging stability, we reduced the LHe flow rate to 100 mbar. Following the experiment, the system undergoes a passive warm-up phase (Stage V).

**Figure S5a** presents the temperature profiles at LHe pressures of 700 mbar and 300 mbar. In both cases, an inflection point is observed at 25 K, marking the transition between stages II and III. However, at a higher pressure, the cooling process proceeds more rapidly than at lower pressure. At 700 mbar, the temperature dropped to 25 K in 132 seconds, followed by an additional 530 seconds to reach the base temperature of 7.1 K. In contrast, at 300 mbar, it took approximately 427 seconds to reach 25 K and 500 seconds to reach the base temperature of 7.9 K. These findings confirm that increased LHe pressure not only accelerates the cooling rate but also enables the system to attain lower base temperatures. **Figure S5b** plots the first derivative of the temperature as a function of time, which provides quantitative insight into the cooling dynamics under different pressure conditions. The most pronounced rate of temperature change is observed during the rapid cooling stage (Stage II), with the maximum derivative reaching –9.24 K/s at 700 mbar and –6.94 K/s at 300 mbar. By averaging the rate of temperature change over stage II, we determined the mean cooling rates to be approximately 2.0 K/s at 700 mbar and 0.6 K/s at 300 mbar. Upon transitioning to Stage III, the temperature change converges to approximately 0.034 K/s for both conditions.

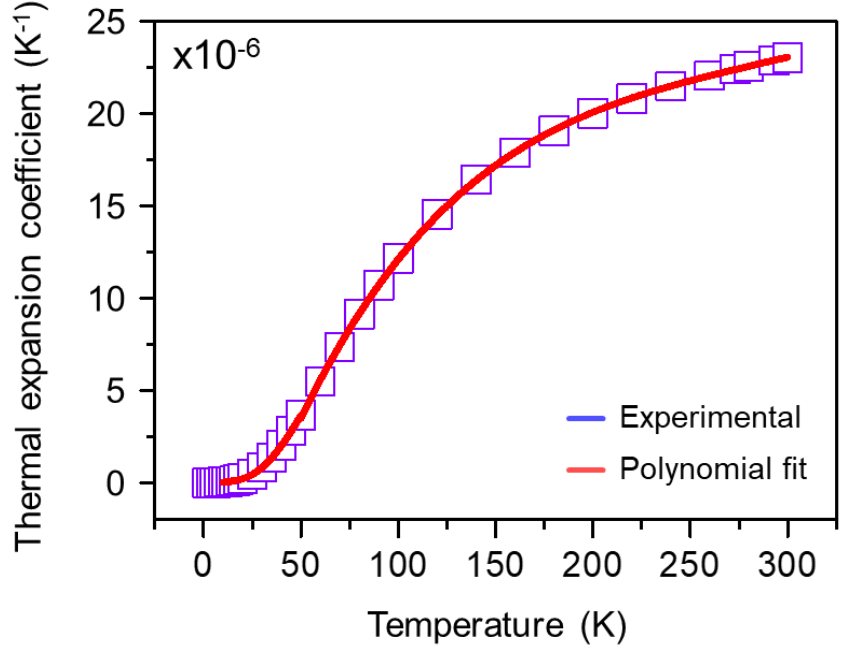

**Figure S6**. Temperature-dependent thermal expansion coefficient ($\alpha_l$) of aluminum, derived from prior study [1]. Due to the rapid variation of $\alpha_l$ at low temperatures, the data were divided into two temperature intervals (10–50 K and 50–300 K), and each segment was fitted using a fourth-order polynomial function. The corresponding fitting constants are provided in **Table S1**. These constants were then integrated to compute the function $f(T) = \int_{T0}^{T} \alpha_l(T)\, dT$, which was subsequently applied to determine the bulk plasmon energy $E_p(T)$.

**Table S1**. Fourth-order polynomial fitting constants for the $\alpha_l(T)$ in the ranges 10–50 K and 50–300 K.

| $\alpha_l(T) = \alpha_0 + \alpha_1(T-T_0) + \alpha_2(T-T_0)^2 + \alpha_3(T-T_0)^3 + \alpha_4(T-T_0)^4$ | | |
|---|---|---|
| Coefficients | 10 K ≤ T ≤ 50K | 50 K ≤ T ≤ 300K |
| $a_0$ | $(3.660 \times 10^{-6}) \pm (2.357 \times 10^{-9})$ | $(2.305 \times 10^{-5}) \pm (6.233 \times 10^{-8})$ |
| $a_1$ | $(1.726 \times 10^{-7}) \pm (9.135 \times 10^{-10})$ | $(2.474 \times 10^{-8}) \pm (4.081 \times 10^{-9})$ |
| $a_2$ | $(0.622 \times 10^{-10}) \pm (9.940 \times 10^{-11})$ | $(9.560 \times 10^{-12}) \pm (7.019 \times 10^{-11})$ |
| $a_3$ | $(-1.011 \times 10^{-10}) \pm (3.754 \times 10^{-12})$ | $(4.158 \times 10^{-13}) \pm (4.270 \times 10^{-13})$ |
| $a_4$ | $(-1.286 \times 10^{-12}) \pm (4.574 \times 10^{-14})$ | $(-1.909 \times 10^{-15}) \pm (8.436 \times 10^{-16})$ |

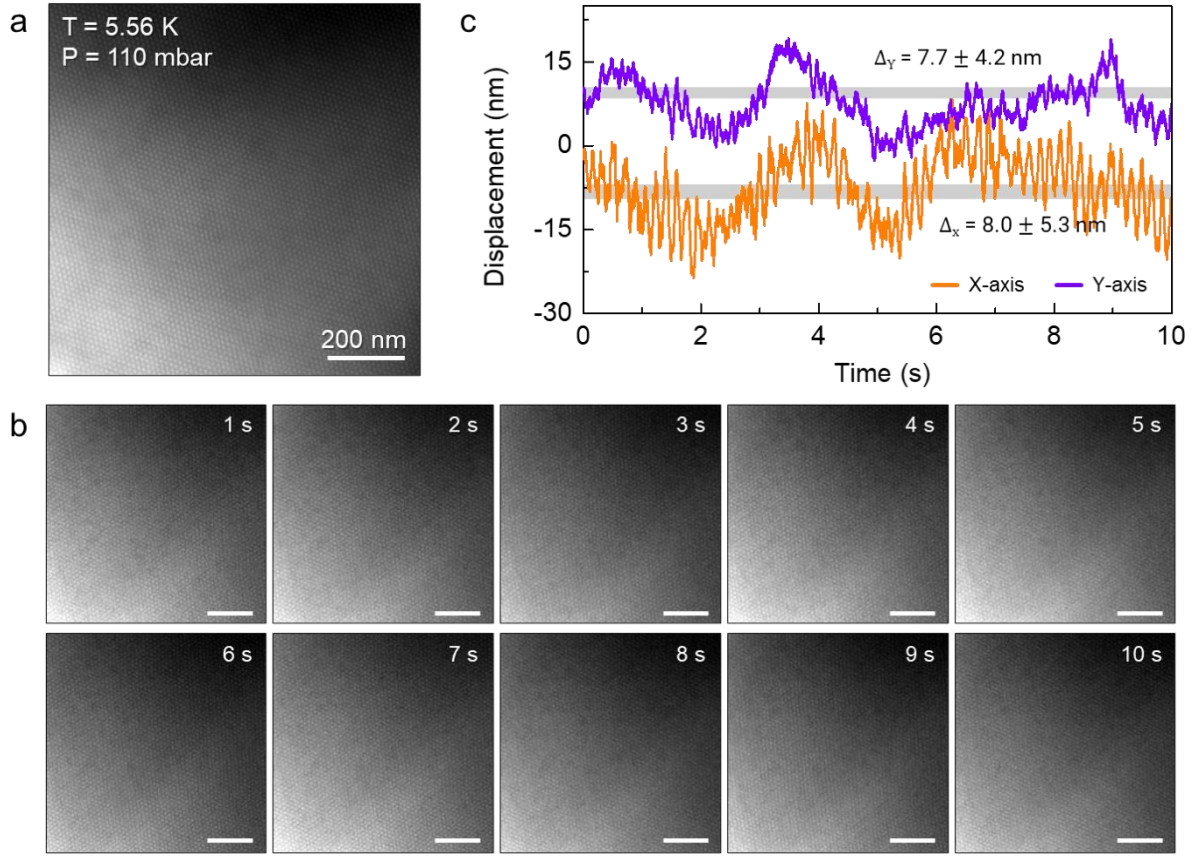

**Figure S7**. **Quantitative assessment of mechanical perturbations during cryogenic Lorentz transmission electron microscopy imaging.** (a) Lorentz micrograph of the hexagonal skyrmion lattice emergent in a MnSi lamellae with defocus $\Delta f = -66$ μm at a temperature of 5.56 K under dewar pressure of 110 mbar. (b) Temporal sequence of micrographs acquired with an exposure time of 0.00177 s over a 10-second acquisition interval. The scale bars are 200 nm. (c) Positional displacement analysis derived from the temporal series, revealing mechanical oscillations of $\Delta_X = 0.80 \pm 0.53$ nm/s and $\Delta_Y = 0.77 \pm 0.42$ nm/s.

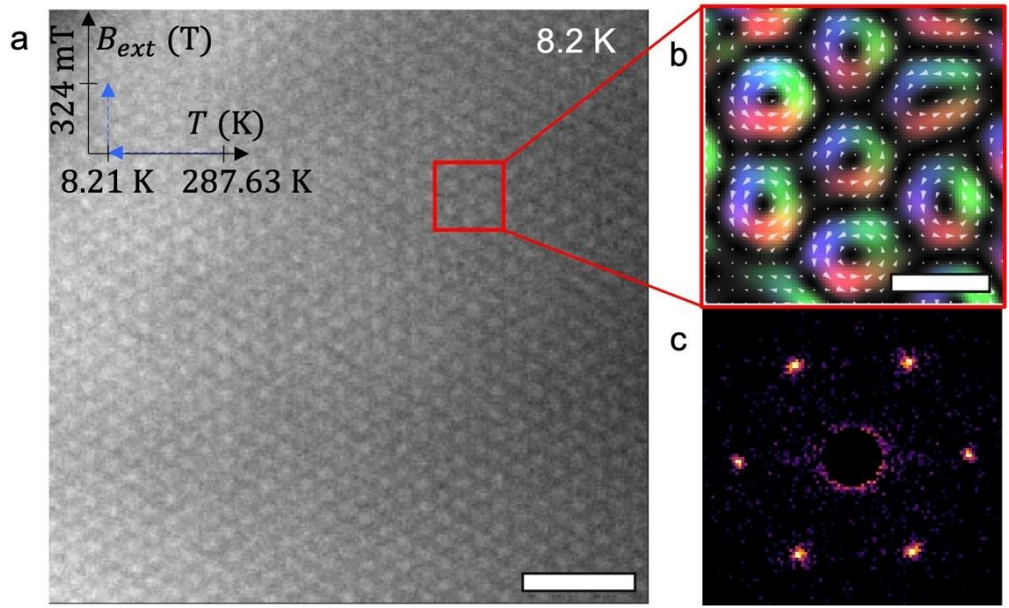

**Figure S8. Magnetic real-space imaging of nanomagnetic textures in MnSi and computed discrete Fourier transform.** (a) Lorentz transmission electron micrograph of the hexagonal skyrmion lattice emergent in a MnSi lamellae with defocus $\Delta f = -33$ μm, 0.98 s exposure time, and applied external magnetic field $\mu_0 H_{ext} = 324$ mT. Scale bar 100 nm. (b) In-plane magnetic induction of the field of view indicated by a red square in (a) reconstructed using the single image transport-of-intensity equation (SITIE). Scale bar 20 nm. (c) Discrete Fourier transform of (a), showing the six-fold symmetry of the skyrmion lattice state. Note that the central region of radius 10 pixels is removed for improved display contrast.